\documentclass[11pt,twoside]{article}
\input{epsf.sty}
\usepackage{mathrsfs}
\usepackage{amsmath}
\usepackage{amssymb}
\usepackage{graphicx}
\usepackage{subfigure}
\newtheorem{proposition}{Proposition}

\title{ Multi-component generalization of Camassa-Holm equation}
\author{Baoqiang Xia$^{1}$\footnote{E-mail address: xiabaoqiang@126.com},
~~Zhijun Qiao$^{2}$\footnote{E-mail address: qiao@utpa.edu}
\\
$^1$School of Mathematics and Statistics, Jiangsu Normal
University,\\
 Xuzhou, Jiangsu 221116, P. R. China
\\ $^{2}$Department of Mathematics, University of Texas-Pan American, \\Edinburg, Texas 78541, USA}

\date{}
\begin{document}
\maketitle
\begin{abstract}
In this paper, we propose a multi-component system of Camassa-Holm equation, denoted by CH($N$,$H$) with $2N$ components and an arbitrary smooth function $H$. This system is shown to admit Lax pair
and infinitely many conservation laws. We particularly study the case of $N=2$ and derive the bi-Hamiltonian structures and peaked soliton (peakon) solutions for some examples.

\noindent {\bf Keywords:}\quad Peakon, Camassa-Holm type equations, Bi-Hamiltonian structure, Lax pair.

\noindent{\bf PACS:}\quad 02.30.Ik, 04.20.Jb.
\end{abstract}

\section{ Introduction}
In 1993, Camassa and Holm derived the well-known Camassa-Holm (CH) equation \cite{CH}
\begin{eqnarray}
m_t+2m u_x+m_xu=0, \quad m=u-u_{xx}+k,
\label{CH}
\end{eqnarray}
(with $k$ being an arbitrary constant) with the aid of an asymptotic approximation to the Hamiltonian of the Green-Naghdi equations.
Since the work of Camassa and Holm \cite{CH}, more diverse studies on this equation have remarkably been developed \cite{CH2}-\cite{CGI}.
The most interesting feature of the CH equation (\ref{CH}) is that it admits peakon solutions in the case $k=0$.
The stability and interaction of peakons were discussed in several references \cite{CS1}-\cite{JR}.
In addition to the CH equation, other similar integrable models with peakon solutions were found \cite{DP1,DP2}.
Recently, there are two integrable peakon equations found with
cubic nonlinearity. They are the following cubic equation \cite{OR,Fo,Fu,Q1}
\begin{eqnarray}
 m_t+\frac{1}{2}\left[ m(u^2-u^2_x)\right]_x=0, \quad  m=u-u_{xx},\label{mCH}
\end{eqnarray}
and the Novikov's equation \cite{NV1,HW1}
\begin{eqnarray}
 m_t=u^2m_x+3uu_xm, \quad  m=u-u_{xx}.\label{cCHN}
\end{eqnarray}
There is also much attention in studying integrable multi-component peakon equations.
For example, in \cite{HI}, the authors proposed a multi-component generalization of the CH equation, and
in \cite{QSY}, multi-component extensions of the cubic nonlinear equation (\ref{mCH}) were studied.

In this paper, we propose the following multi-component system
\begin{eqnarray}
\left\{\begin{split}
m_{j,t}=&(m_jH)_x+m_jH+\frac{1}{(N+1)^2}\sum_{i=1}^{N}[m_{i}(u_j-u_{j,x})(v_{i}+v_{i,x})+m_j(u_i-u_{i,x})(v_{i}+v_{i,x})],
\\
n_{j,t}=&(n_jH)_x-n_jH-\frac{1}{(N+1)^2}\sum_{i=1}^{N}[n_{i}(u_i-u_{i,x})(v_{j}+v_{j,x})+n_j(u_i-u_{i,x})(v_{i}+v_{i,x})],
\\
m_{j}=&u_{j}-u_{j,xx}, \quad n_{j}=v_{j}-v_{j,xx}, \quad 1\leq j\leq N,
\end{split}\right. \label{meq}
\end{eqnarray}
where $H$ is an arbitrary smooth function of $u_j$, $v_j$, $1\leq j\leq N$, and their derivatives.
For $N=1$, this system is reduced to the standard CH equation (\ref{CH}) as $v_{1}=2$, $H=-u_1$ and to the cubic nonlinear CH equation (\ref{mCH}) as $u_1=v_{1}$, $H=-\frac{1}{2}(u_{1}^{2}-u_{1,x}^{2})$.
Therefore, it is a kind of multi-component combination of the CH equation (\ref{CH}) and the cubic nonlinear CH equation (\ref{mCH}).
The system (\ref{meq}) contains an arbitrary function $H$, thus it is actually a large class of multi-component equations.
We remark that, very recently, Li, Liu and Popowicz proposed a four-component peakon equation which also contains an arbitrary function \cite{LLP}.
They derived the Lax pair and infinite conservation laws for their four-component equation,
and presented a bi-Hamiltonian structure for their equation in the case that the arbitrary function is taken to be zero.
In this paper, we show that the multi-component system (\ref{meq}) admits Lax representation and infinitely many conservation laws.
Due to the presence of the arbitrary function $H$, we do not expect the system (\ref{meq}) is bi-Hamiltonian integrable system in general.
However, we demonstrate that for some special choices of $H$ we may find the corresponding bi-Hamiltonian structures.
As examples, we derive the peakon solutions of this system in the case $N=2$.
In particular, we obtain a new integrable model which admits stationary peakon solutions.

The whole paper is organized as follows. In section 2, the Lax
pair and conservation laws of equation (\ref{meq}) are presented.
In section 3, the Hamiltonian structures and peakon solutions of equation (\ref{meq}) in the case $N=2$ are discussed.
Some conclusions and open problems are addressed in section 4.

\section{Lax pair and conservation laws}

We first introduce the $N$-component vector potentials $\vec{u}$, $\vec{v}$ and $\vec{m}$, $\vec{n}$ as
\begin{eqnarray}
\begin{split}
\vec{u}=&(u_1,u_2,\cdots,u_N),
\quad
\vec{v}=(v_1,v_2,\cdots,v_N),
\\
\vec{m}=&\vec{u}-\vec{u}_{xx}, \quad \vec{n}=\vec{v}-\vec{v}_{xx}.
\end{split}
\label{um}
\end{eqnarray}
Using this notation, equation (\ref{meq}) is expressed in the following vector form
\begin{eqnarray}
\left\{\begin{split}
\vec{m}_{t}=&(\vec{m}H)_x+\vec{m}H+\frac{1}{(N+1)^2}[\vec{m}(\vec{v}+\vec{v}_x)^T(\vec{u}-\vec{u}_x)+(\vec{u}-\vec{u}_x)(\vec{v}+\vec{v}_x)^T\vec{m}],
\\
\vec{n}_{t}=&(\vec{n}H)_x-\vec{n}H-\frac{1}{(N+1)^2}[\vec{n}(\vec{u}-\vec{u}_x)^T(\vec{v}+\vec{v}_x)+(\vec{v}+\vec{v}_x)(\vec{u}-\vec{u}_x)^T\vec{n}],
\\
\vec{m}=&\vec{u}-\vec{u}_{xx}, \quad \vec{n}=\vec{v}-\vec{v}_{xx},
\end{split}\right.
\label{veq}
\end{eqnarray}
where the symbol $T$ denotes the transpose of a vector.

Let us introduce a pair of $(N+1)\times(N+1)$ matrix spectral problems
\begin{eqnarray}
\phi_x=U\phi,\quad \phi_t=V\phi,
\label{LP}
\end{eqnarray}
with
\begin{eqnarray}
\begin{split}
\phi&=(\phi_1,\phi_{21},\cdots,\phi_{2N})^T,\\
U&=\frac{1}{N+1}\left( \begin{array}{cc} -N & \lambda \vec{m}\\
 \lambda  \vec{n}^T &  I_N \\ \end{array} \right),
\\
V&=\frac{1}{N+1}\left( \begin{array}{cc} -N\lambda^{-2}+\frac{1}{N+1}(\vec{u}-\vec{u}_x)(\vec{v}+\vec{v}_x)^T &
\lambda^{-1}(\vec{u}-\vec{u}_x)+\lambda \vec{m}H
\\ \lambda^{-1}(\vec{v}+\vec{v}_x)^T+\lambda \vec{n}^TH & \lambda^{-2}I_N-\frac{1}{N+1}(\vec{v}+\vec{v}_x)^T(\vec{u}-\vec{u}_x)\\ \end{array} \right),
\label{UV}
\end{split}
\end{eqnarray}
where $\lambda$ is a spectral parameter, $I_N$ is the $N\times N$ identity matrix,
$\vec{u}$, $\vec{v}$, $\vec{m}$ and $\vec{n}$ are the vector potentials shown in (\ref{um}).

\begin{proposition}
(\ref{LP}) provides the Lax pair for the multi-component system (\ref{meq}).
\end{proposition}
{\bf Proof} \quad
It is easy to check that the compatibility condition of (\ref{LP}) generates
\begin{eqnarray}
U_t-V_x+[U,V]=0.\label{cc}
\end{eqnarray}
From (\ref{UV}), we have
\begin{eqnarray}
\begin{split}
U_t&=\frac{1}{N+1}\left( \begin{array}{cc} 0 & \lambda \vec{m}_t\\
 \lambda  \vec{n}_t^T &  0 \\ \end{array} \right),
\\
V_x&=\frac{1}{N+1}\left( \begin{array}{cc} \frac{1}{N+1}[\vec{m}(\vec{v}+\vec{v}_x)^T-(\vec{u}-\vec{u}_x)\vec{n}^T] &
\lambda^{-1}(\vec{u}_x-\vec{u}_{xx})+\lambda (\vec{m}H)_x
\\ \lambda^{-1}(\vec{v}_x+\vec{v}_{xx})^T+\lambda (\vec{n}^TH)_x & \frac{1}{N+1}[\vec{n}^T(\vec{u}-\vec{u}_x)-(\vec{v}+\vec{v}_x)^T\vec{m}]\\ \end{array} \right),
\label{UVd}
\end{split}
\end{eqnarray}
and
\begin{eqnarray}
\begin{split}
[U,V]=UV-VU=\frac{1}{(N+1)^2}\left( \begin{array}{cc} \Gamma_{11} &
\Gamma_{12}
\\ \Gamma_{21} & \Gamma_{22}\\ \end{array} \right),
\label{UVc}
\end{split}
\end{eqnarray}
where
\begin{eqnarray*}
\begin{split}
\Gamma_{11}&= \vec{m}(\vec{v}+\vec{v}_x)^T-(\vec{u}-\vec{u}_x)\vec{n}^T,
\\
\Gamma_{12}&=(N+1)[\lambda^{-1}(\vec{u}_x-\vec{u}_{xx})-\lambda\vec{m}H]-
\frac{\lambda}{N+1}[\vec{m}(\vec{v}+\vec{v}_x)^T(\vec{u}-\vec{u}_x)+(\vec{u}-\vec{u}_x)(\vec{v}+\vec{v}_x)^T\vec{m}],
\\
\Gamma_{21}&=(N+1)[\lambda^{-1}(\vec{v}_x+\vec{v}_{xx})^T+\lambda\vec{n}^TH]+
\frac{\lambda}{N+1}[\vec{n}^T(\vec{u}-\vec{u}_x)(\vec{v}+\vec{v}_x)^T+(\vec{v}+\vec{v}_x)^T(\vec{u}-\vec{u}_x)\vec{n}^T],
\\
\Gamma_{22}&= \vec{n}^T(\vec{u}-\vec{u}_x)-(\vec{v}+\vec{v}_x)^T\vec{m}.
\label{gma}
\end{split}
\end{eqnarray*}
We remark that (\ref{UVc}) is written in the form of block matrix. As shown above, the element $\Gamma_{11}$ is a scalar function, the element $\Gamma_{12}$ is a $N$-component row vector function, the element $\Gamma_{21}$ is a $N$-component column vector function, and the element $\Gamma_{22}$ is a $N \times N$ matrix function.

Substituting the expressions (\ref{UVd}) and (\ref{UVc}) into (\ref{cc}), we find that (\ref{cc}) gives rise to
\begin{eqnarray}
\left\{\begin{split}
\vec{m}_{t}=&(\vec{m}H)_x+\vec{m}H+\frac{1}{(N+1)^2}[\vec{m}(\vec{v}+\vec{v}_x)^T(\vec{u}-\vec{u}_x)+(\vec{u}-\vec{u}_x)(\vec{v}+\vec{v}_x)^T\vec{m}],
\\
\vec{n}^T_{t}=&(\vec{n}^TH)_x-\vec{n}^TH-\frac{1}{(N+1)^2}[\vec{n}^T(\vec{u}-\vec{u}_x)(\vec{v}+\vec{v}_x)^T+(\vec{v}+\vec{v}_x)^T(\vec{u}-\vec{u}_x)\vec{n}^T],
\\
\vec{m}=&\vec{u}-\vec{u}_{xx}, \quad \vec{n}^T=\vec{v}^T-\vec{v}_{xx}^T,
\end{split}\right.
\label{veq2}
\end{eqnarray}
which is nothing but the vector equation (\ref{veq}). Hence, (\ref{LP}) exactly gives the Lax pair of multi-component equation (\ref{meq}).

Now let us construct the conservation laws of equation (\ref{meq}).
We write the spacial part of the spectral problems (\ref{LP}) as
\begin{eqnarray}
\left\{\begin{array}{l}
\phi_{1,x}=\frac{1}{N+1}(-N\phi_{1}+\lambda\sum_{i=1}^Nm_i\phi_{2i}),
\\
\phi_{2j,x}=\frac{1}{N+1}\left(\lambda n_j\phi_{1}+\phi_{2j}\right),
\end{array}\right.
1\leq j\leq N.
\label{slp}
\end{eqnarray}
Let $\Omega_j=\frac{\phi_{2j}}{\phi_{1}}$, $1\leq j\leq N$, we obtain the following system of Riccati equations
\begin{eqnarray}
\Omega_{j,x}=\frac{1}{N+1}\left[\lambda n_j+(N+1)\Omega_{j}-\lambda\Omega_{j}\sum_{i=1}^Nm_i\Omega_{i}\right], ~~1\leq j\leq N.
\label{ric}
\end{eqnarray}
Making use of the relation $(\ln\phi_1)_{xt}=(\ln\phi_1)_{tx}$ and (\ref{LP}), we arrive at the conservation law
\begin{eqnarray}
\left(\sum_{i=1}^Nm_i\Omega_{i}\right)_t=
\left(\lambda^{-2}\sum_{i=1}^N(u_i-u_{i,x})\Omega_{i}+\frac{1}{N+1}\lambda^{-1}\sum_{i=1}^N(u_i-u_{i,x})(v_{i}+v_{i,x})+H\sum_{i=1}^Nm_i\Omega_{i}\right)_x.
\label{CL}
\end{eqnarray}

Equation (\ref{CL}) means that $\sum_{i=1}^Nm_i\Omega_{i}$ is a generating function of the conserved densities. To derive the explicit forms of conserved densities, we expand $\Omega_j$ into the negative power series of $\lambda$ as
\begin{equation}
\Omega_j=\sum_{k=0}^{\infty}\omega_{jk}\lambda^{-k}, \quad 1\leq j\leq N.\label{oe1}
\end{equation}
Substituting (\ref{oe1}) into the Riccati system (\ref{ric}) and equating the coefficients of powers of $\lambda$, we obtain
\begin{eqnarray}
\begin{split}
\omega_{j0}&=n_j\left(\sum_{i=1}^Nm_in_i\right)^{-\frac{1}{2}},
\\
\omega_{j1}&=(N+1)\left[\omega_{j0}-\omega_{j0,x}-
\frac{1}{2}n_j\left(\sum_{i=1}^Nm_i(\omega_{i0}-\omega_{i0,x})\right)\left(\sum_{i=1}^Nm_in_i\right)^{-1}\right]\left(\sum_{i=1}^Nm_in_i\right)^{-\frac{1}{2}},
\end{split}
\label{wj1}
\end{eqnarray}
and the recursion relations for $\omega_{j(k+1)}$, $k\geq 1$,
\begin{eqnarray}
\begin{split}
\omega_{j(k+1)}&=(N+1)\left[\omega_{jk}-\omega_{jk,x}-\frac{1}{2}n_j\left(\sum_{i=1}^Nm_i(\omega_{ik}-\omega_{ik,x})\right)
\left(\sum_{i=1}^Nm_in_i\right)^{-1}\right]\left(\sum_{i=1}^Nm_in_i\right)^{-\frac{1}{2}}.
\end{split}
\label{wj}
\end{eqnarray}

Inserting (\ref{oe1}), (\ref{wj1}) and (\ref{wj}) into (\ref{CL}), we finally obtain the following infinitely many conserved densities $\rho_j$
and the associated fluxes $F_j$:
\begin{eqnarray}
\begin{split}
\rho_{0}&=\sum_{i=1}^Nm_i\omega_{i0}=\left(\sum_{i=1}^Nm_in_i\right)^{\frac{1}{2}}, ~~F_0=H\sum_{i=1}^Nm_i\omega_{i0}=H\left(\sum_{i=1}^Nm_in_i\right)^{\frac{1}{2}},
\\
\rho_{1}&=\sum_{i=1}^Nm_i\omega_{i1}, ~~ F_1=\frac{1}{N+1}\sum_{i=1}^N(u_i-u_{i,x})(v_{i}+v_{i,x})+H\sum_{i=1}^Nm_i\omega_{i1},
\\
\rho_{2}&=\sum_{i=1}^Nm_i\omega_{i2}, ~~ F_2=\sum_{i=1}^N(u_i-u_{i,x})\omega_{i0}+H\sum_{i=1}^Nm_i\omega_{i2},
\\
\rho_{j}&=\sum_{i=1}^Nm_i\omega_{ij}, ~~F_{j}=\sum_{i=1}^N(u_i-u_{i,x})\omega_{i(j-2)}+H\sum_{i=1}^Nm_i\omega_{ij},\quad j\geq 3,
\end{split}
\label{rjj}
\end{eqnarray}
where $\omega_{ij}$, $1\leq i\leq N$, $j\geq 0$ is given by (\ref{wj1}) and (\ref{wj}).

\vspace*{0.2cm}
{\bf Remark 1.} The $2N$-component system (\ref{meq}) containing an arbitrary function $H$ does possess Lax representation and infinitely many conservation laws. Such a system is interesting since different choices of $H$ lead to different peakon equations (see examples in the following section). Let us look back why an arbitrary smooth function may be involved in system (\ref{meq}). System (\ref{meq}) is produced by the compatibility condition (\ref{cc}) of the spectral problems (\ref{LP}) where such an arbitrary function is included in $V$ part (see (\ref{UV})). The appearance of this arbitrary function can be explained as that the Lax equation is an over determined system by choosing the appropriate $V$ to match $U$.

\section{Examples for $N=2$}

For $N=1$, equation (\ref{meq}) becomes
\begin{eqnarray}
\left\{\begin{split}
m_{1,t}=&(m_1H)_x+m_1H+\frac{1}{2}m_{1}(u_1-u_{1,x})(v_{1}+v_{1,x}),
\\
n_{1,t}=&(n_1H)_x-n_1H-\frac{1}{2}n_{1}(u_1-u_{1,x})(v_{1}+v_{1,x}),
\\
m_{1}=&u_{1}-u_{1,xx}, \quad n_{1}=v_{1}-v_{1,xx},
\end{split}\right. \label{teq0}
\end{eqnarray}
where $H$ is an arbitrary smooth function of $u_1$, $v_1$, and their derivatives.
This system is exactly the synthetical two-component peakon model we proposed in \cite{XQZ}, where one may see the details of the Lax pair, bi-Hamiltonian structures and peakon solutions of this model.

Let us study the case of $N=2$. In this case, equation (\ref{meq}) is cast into the following four-component model
\begin{eqnarray}
\left\{
\begin{split}
m_{1,t}=&(m_1H)_x+m_1H
\\&+\frac{1}{9}\{m_{1}[2(u_{1}-u_{1,x})(v_{1}+v_{1,x})+(u_{2}-u_{2,x})(v_{2}+v_{2,x})]+m_{2}(u_{1}-u_{1,x})(v_{2}+v_{2,x})\},
\\
m_{2,t}=&(m_2H)_x+m_2H
\\&+\frac{1}{9}\{m_{1}(u_{2}-u_{2,x})(v_{1}+v_{1,x})+m_{2}[(u_{1}-u_{1,x})(v_{1}+v_{1,x})+2(u_{2}-u_{2,x})(v_{2}+v_{2,x})]\},
\\
n_{1,t}=&(n_1H)_x-n_1H
\\&-\frac{1}{9}\{n_{1}[2(u_{1}-u_{1,x})(v_{1}+v_{1,x})+(u_{2}-u_{2,x})(v_{2}+v_{2,x})]+n_{2}(u_{2}-u_{2,x})(v_{1}+v_{1,x})\},
\\
n_{2,t}=&(n_2H)_x-n_2H
\\&-\frac{1}{9}\{n_{1}(u_{1}-u_{1,x})(v_{2}+v_{2,x})+n_{2}[(u_{1}-u_{1,x})(v_{1}+v_{1,x})+2(u_{2}-u_{2,x})(v_{2}+v_{2,x})]\},
\\ m_{1}=&u_{1}-u_{1,xx},\quad  m_{2}=u_{2}-u_{2,xx}, \quad n_{1}=v_{1}-v_{1,xx},\quad  n_{2}=v_{2}-v_{2,xx},
\end{split}
\right.
\label{teq1}
\end{eqnarray}
where $H$ is an arbitrary smooth function of $u_1$,  $u_2$,  $v_1$, $v_2$, and their derivatives. This system admits the following $3\times 3$ Lax pair
\begin{eqnarray}
U=\frac{1}{3}\left( \begin{array}{ccc} -2 & \lambda m_1 & \lambda m_2\\
 \lambda n_1 &  1 & 0
 \\ \lambda n_2 & 0 & 1 \\ \end{array} \right),
 \quad
 V=\frac{1}{3}\left( \begin{array}{ccc} V_{11} & V_{12} & V_{13} \\
V_{21} & V_{22} & V_{23} \\ V_{31} & V_{32} & V_{33}\\ \end{array} \right),
\label{UV1}
\end{eqnarray}
where
\begin{eqnarray}
\begin{split}
V_{11}&=-2\lambda^{-2}+\frac{1}{3}[(u_{1}-u_{1,x})(v_{1}+v_{1,x})+(u_{2}-u_{2,x})(v_{2}+v_{2,x})],
\\
V_{12}&=\lambda^{-1}(u_1-u_{1,x})+\lambda m_1H, ~~V_{13}=\lambda^{-1}(u_2-u_{2,x})+\lambda m_2H,
\\
V_{21}&=\lambda^{-1}(v_1+v_{1,x})+\lambda n_1H, ~~V_{22}=\lambda^{-2}-\frac{1}{3}(u_{1}-u_{1,x})(v_{1}+v_{1,x}),
\\
V_{23}&=-\frac{1}{3}(u_{2}-u_{2,x})(v_{1}+v_{1,x}), ~~V_{31}=\lambda^{-1}(v_2+v_{2,x})+\lambda n_2H,
\\
V_{32}&=-\frac{1}{3}(u_{1}-u_{1,x})(v_{2}+v_{2,x}), ~~V_{33}=\lambda^{-2}-\frac{1}{3}(u_{2}-u_{2,x})(v_{2}+v_{2,x}).
\label{v}
\end{split}
\end{eqnarray}
Due to the appearance of arbitrary function $H$, we do not expect that (\ref{teq1}) is bi-Hamiltonian in general.
But we find that it is possible to figure out the bi-Hamiltonian structures for some special cases, which we will show in the following examples.

\subsection*{Example 1.~~A new integrable model with stationary peakon solutions}
Taking $H=0$, equation (\ref{teq1}) becomes the following system
\begin{eqnarray}
\left\{\begin{array}{l}
m_{1,t}=\frac{1}{9}\{m_{1}[2(u_{1}-u_{1,x})(v_{1}+v_{1,x})+(u_{2}-u_{2,x})(v_{2}+v_{2,x})]+m_{2}(u_{1}-u_{1,x})(v_{2}+v_{2,x})\},
\\
m_{2,t}=\frac{1}{9}\{m_{1}(u_{2}-u_{2,x})(v_{1}+v_{1,x})+m_{2}[(u_{1}-u_{1,x})(v_{1}+v_{1,x})+2(u_{2}-u_{2,x})(v_{2}+v_{2,x})]\},
\\
n_{1,t}=-\frac{1}{9}\{n_{1}[2(u_{1}-u_{1,x})(v_{1}+v_{1,x})+(u_{2}-u_{2,x})(v_{2}+v_{2,x})]+n_{2}(u_{2}-u_{2,x})(v_{1}+v_{1,x})\},
\\
n_{2,t}=-\frac{1}{9}\{n_{1}(u_{1}-u_{1,x})(v_{2}+v_{2,x})+n_{2}[(u_{1}-u_{1,x})(v_{1}+v_{1,x})+2(u_{2}-u_{2,x})(v_{2}+v_{2,x})]\},
\\ m_{1}=u_{1}-u_{1,xx},\quad  m_{2}=u_{2}-u_{2,xx},
\quad n_{1}=v_{1}-v_{1,xx},\quad  n_{2}=v_{2}-v_{2,xx}.
\end{array}\right. \label{teq2}
\end{eqnarray}
Let us introduce a Hamiltonian pair
\begin{eqnarray}
\begin{split}
J&=\left( \begin{array}
{cccc} 0 &  0 & \partial+1 & 0 \\  0 &  0 & 0 & \partial+1 \\ \partial-1 &  0 & 0 &0 \\ 0 & \partial-1 & 0 &0 \\
\end{array} \right),
~~K=\frac{1}{9}\left( \begin{array}
{cccc} K_{11} & K_{12}  & K_{13} & K_{14} \\
K_{21} &  K_{22}  & K_{23} & K_{24} \\
K_{31} &  K_{32} & K_{33} & K_{34} \\
K_{41} &  K_{42} & K_{43} & K_{44} \\
\end{array} \right),
\label{JK1}
\end{split}
\end{eqnarray}
where
\begin{eqnarray}
\begin{split}
K_{11}=& -2m_1\partial^{-1}m_1, ~~K_{12}=-m_2\partial^{-1}m_1-m_1\partial^{-1}m_2,
\\ K_{13}=& 2m_1\partial^{-1}n_1+m_2\partial^{-1}n_2, ~~K_{14}=m_1\partial^{-1}n_2,
\\ K_{21}=&-K_{12}^{\ast}=-m_1\partial^{-1}m_2-m_2\partial^{-1}m_1, ~~K_{22}=-2m_2\partial^{-1}m_2,
\\ K_{23}=& m_{2}\partial^{-1} n_{1}, ~~K_{24}=m_1\partial^{-1}n_1+2m_2\partial^{-1}n_2,
\\ K_{31}=&-K_{13}^{\ast}=2n_1\partial^{-1}m_1+n_2\partial^{-1}m_2, ~~K_{32}=-K_{23}^{\ast}=n_1\partial^{-1} m_2,
\\ K_{33}=&-2 n_{1}\partial^{-1} n_{1}, ~~K_{34}=-n_{1}\partial^{-1} n_{2}-n_{2}\partial^{-1} n_{1},
\\ K_{41}=&-K_{14}^{\ast}=n_2\partial^{-1} m_1, ~~K_{42}=-K_{24}^{\ast}=n_1\partial^{-1}m_1+2n_2\partial^{-1}m_2,
\\ K_{43}=&-K_{34}^{\ast}=-n_2\partial^{-1}n_1-n_1\partial^{-1}n_2, ~~K_{44}=-2n_2\partial^{-1}n_2.
\end{split}
\label{K1}
\end{eqnarray}
By direct but tedious calculations, we arrive at
\begin{proposition}
Equation (\ref{teq2}) can be rewritten in the following bi-Hamiltonian form
\begin{eqnarray}
\left(m_{1,t}, ~m_{2,t}, ~n_{1,t}, ~n_{2,t}\right)^{T}
=J \left(\frac{\delta H_2}{\delta m_{1}},~\frac{\delta H_2}{\delta  m_{2}},~\frac{\delta H_2}{\delta n_{1}},~\frac{\delta H_2}{\delta n_{2}}\right)^{T}
=K \left(\frac{\delta H_1}{\delta m_{1}},~\frac{\delta H_1}{\delta  m_{2}},~\frac{\delta H_1}{\delta n_{1}},~\frac{\delta H_1}{\delta n_{2}}\right)^{T},
\label{BH}
\end{eqnarray}
where $J$ and $K$ are given by (\ref{JK1}), and
\begin{eqnarray}
\begin{split}
H_1&=\int_{-\infty}^{+\infty}[(u_{1,x}-u_{1})n_{1}+(u_{2,x}-u_{2})n_{2}]dx,
\\ H_2&=\frac{1}{9}\int_{-\infty}^{+\infty}[(u_{1}-u_{1,x})^2(v_{1}+v_{1,x})n_1+(u_{1}-u_{1,x})(u_{2}-u_{2,x})(v_{2}+v_{2,x})n_1
\\&~~~~~~~~~~~~ +(u_{1}-u_{1,x})(u_{2}-u_{2,x})(v_{1}+v_{1,x})n_2+(u_{2}-u_{2,x})^2(v_{2}+v_{2,x})n_2]dx.
\end{split}
\label{H}
\end{eqnarray}
\end{proposition}

Suppose an $N$-peakon solution of (\ref{teq2}) is in the form
\begin{eqnarray}
\begin{split}
u_{1}&=\sum_{j=1}^N p_j(t)e^{-\mid x-q_j(t)\mid}, ~~u_{2}=\sum_{j=1}^N r_j(t)e^{-\mid x-q_j(t)\mid},
\\
v_{1}&=\sum_{j=1}^N s_j(t)e^{-\mid x-q_j(t)\mid}, ~~v_{2}=\sum_{j=1}^N w_j(t)e^{-\mid x-q_j(t)\mid}.
\end{split}
\label{NP2}
\end{eqnarray}
Then, in the distribution sense, one can get
\begin{eqnarray}
\begin{split}
u_{1,x}&=-\sum_{j=1}^N p_jsgn(x-q_j)e^{-\mid x-q_j\mid}, \quad m_{1}=2\sum_{j=1}^N p_j\delta(x-q_j),
\\
u_{2,x}&=-\sum_{j=1}^N r_jsgn(x-q_j)e^{-\mid x-q_j\mid}, \quad m_{2}=2\sum_{j=1}^N r_j\delta(x-q_j),
\\
v_{1,x}&=-\sum_{j=1}^N s_jsgn(x-q_j)e^{-\mid x-q_j\mid}, \quad n_{1}=2\sum_{j=1}^N s_j\delta(x-q_j),
\\
v_{2,x}&=-\sum_{j=1}^N w_jsgn(x-q_j)e^{-\mid x-q_j\mid}, \quad n_{2}=2\sum_{j=1}^N w_j\delta(x-q_j).
\end{split}
\label{Npd}
\end{eqnarray}
Substituting (\ref{NP2}) and (\ref{Npd}) into (\ref{teq2}) and integrating against test functions with compact support, we arrive at the $N$-peakon dynamical system as follows:
\begin{eqnarray}
\begin{split}
q_{j,t}=&0,\\
p_{j,t}=&-\frac{1}{9}\{\frac{2}{3}p_j(p_js_j+r_jw_j)
         \\&-\sum_{i,k=1}^N \left(p_j(2p_is_k+r_iw_k)+r_jp_iw_k\right)\left(sgn(q_j-q_i)-sgn(q_j-q_k)\right)e^{ -\mid q_j-q_i\mid-\mid q_j-q_k\mid}
         \\&+\sum_{i,k=1}^N \left(p_j(2p_is_k+r_iw_k)+r_jp_iw_k\right)\left(sgn(q_j-q_i)sgn(q_j-q_k)-1\right)e^{ -\mid q_j-q_i\mid-\mid q_j-q_k\mid}\},\\
r_{j,t}=&-\frac{1}{9}\{\frac{2}{3}r_j(r_jw_j+p_js_j)
         \\&-\sum_{i,k=1}^N \left(r_j(2r_iw_k+p_is_k)+p_jr_is_k\right)\left(sgn(q_j-q_i)-sgn(q_j-q_k)\right)e^{ -\mid q_j-q_i\mid-\mid q_j-q_k\mid}
         \\&+\sum_{i,k=1}^N \left(r_j(2r_iw_k+p_is_k)+p_jr_is_k\right)\left(sgn(q_j-q_i)sgn(q_j-q_k)-1\right)e^{ -\mid q_j-q_i\mid-\mid q_j-q_k\mid}\},\\
s_{j,t}=&\frac{1}{9}\{\frac{2}{3}s_j(p_js_j+r_jw_j)
         \\&-\sum_{i,k=1}^N \left(s_j(2p_is_k+r_iw_k)+w_jr_is_k\right)\left(sgn(q_j-q_i)-sgn(q_j-q_k)\right)e^{ -\mid q_j-q_i\mid-\mid q_j-q_k\mid}
         \\&+\sum_{i,k=1}^N \left(s_j(2p_is_k+r_iw_k)+w_jr_is_k\right)\left(sgn(q_j-q_i)sgn(q_j-q_k)-1\right)e^{ -\mid q_j-q_i\mid-\mid q_j-q_k\mid}\},\\
w_{j,t}=&\frac{1}{9}\{\frac{2}{3}w_j(p_js_j+r_jw_j)
         \\&-\sum_{i,k=1}^N \left(w_j(2r_iw_k+p_is_k)+s_jp_iw_k\right)\left(sgn(q_j-q_i)-sgn(q_j-q_k)\right)e^{ -\mid q_j-q_i\mid-\mid q_j-q_k\mid}
         \\&+\sum_{i,k=1}^N \left(w_j(2r_iw_k+p_is_k)+s_jp_iw_k\right)\left(sgn(q_j-q_i)sgn(q_j-q_k)-1\right)e^{ -\mid q_j-q_i\mid-\mid q_j-q_k\mid}\}.
\end{split}
\label{dNcp2}
\end{eqnarray}

\begin{figure}
\begin{minipage}[t]{0.5\linewidth}
\centering
\includegraphics[width=2.2in]{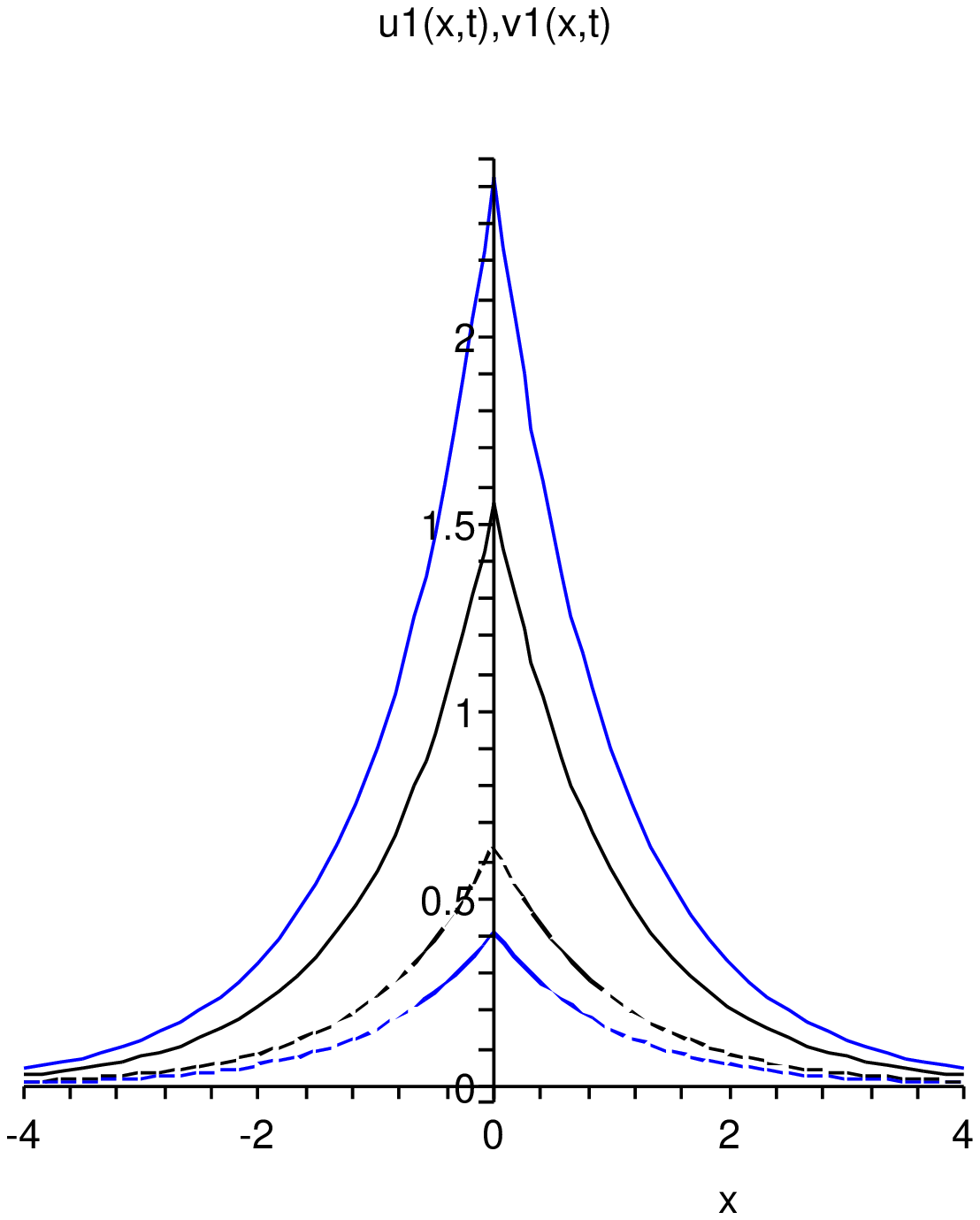}
\caption{\small{ The stationary single-peakon solution of the potentials $u_{1}(x,t)$ and $v_{1}(x,t)$ given by (\ref{ocpnp2}) with $C_1=0$, $A_2=A_4=1$ and $A_3=2$. Solid line: $u_{1}(x,t)$; Dashed line: $v_{1}(x,t)$; Black: $t=1$; Blue: $t=2$. }}
\label{F1}
\end{minipage}
\hspace{2.0ex}
\begin{minipage}[t]{0.5\linewidth}
\centering
\includegraphics[width=2.2in]{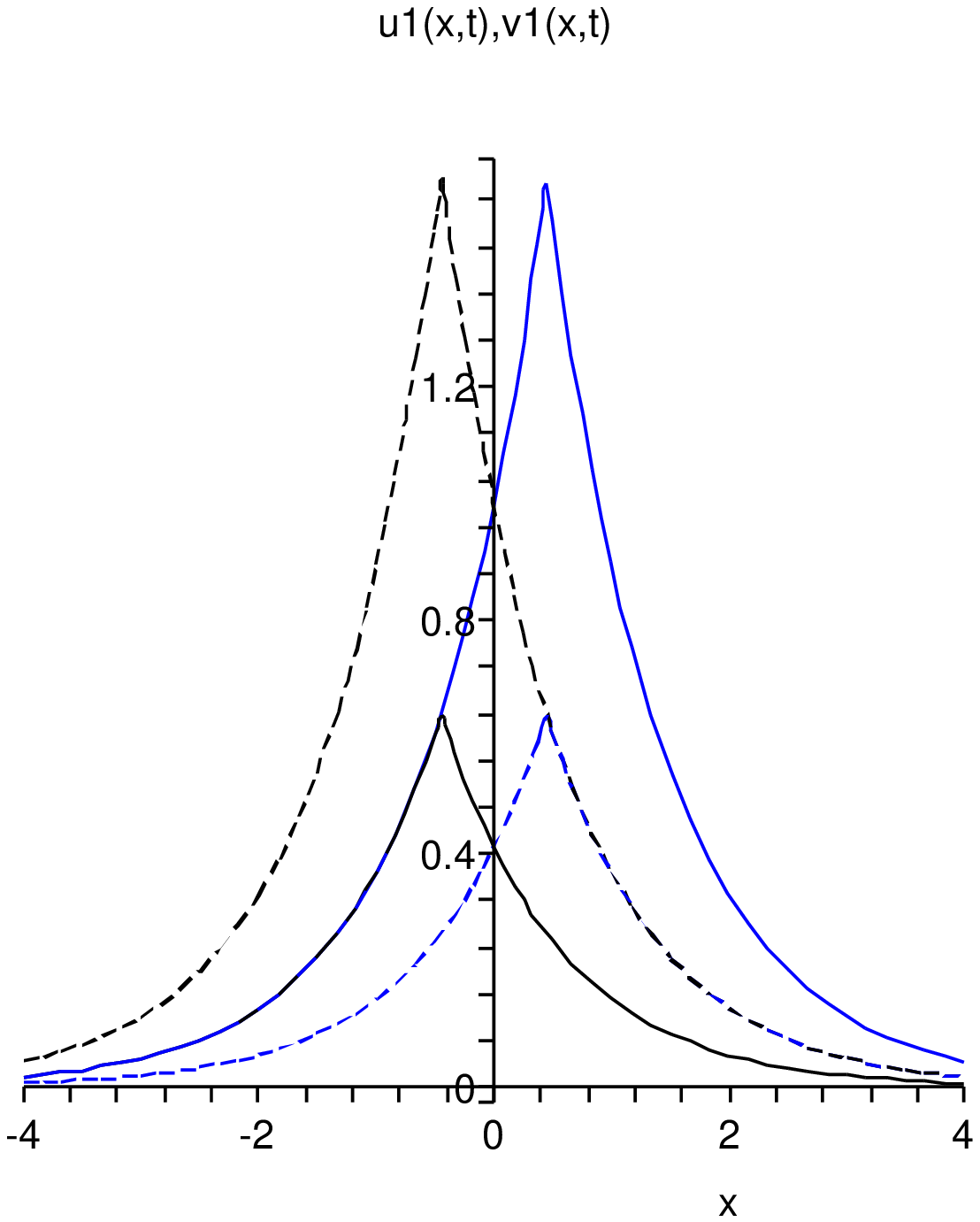}
\caption{\small{ The single-peakon solution of the potentials $u_{1}(x,t)$ and $v_{1}(x,t)$ given by (\ref{N2s}) with $A_4=0$, $A_2=A_5=1$ and $A_3=2$. Solid line: $u_{1}(x,t)$; Dashed line: $v_{1}(x,t)$; Black: $t=-2$; Blue: $t=2$. }}
\label{F2}
\end{minipage}
\end{figure}

The formula $q_{j,t}=0$ in (\ref{dNcp2}) implies that the peakon position is stationary and the solution is in the form of separation of variables.
Especially, for $N=1$, (\ref{dNcp2}) becomes
\begin{eqnarray}
\left\{\begin{array}{l}
q_{1,t}=0,
\\
p_{1,t}=\frac{4}{27}p_1(p_1s_1+r_1w_1),
\\
r_{1,t}=\frac{4}{27}r_1(p_1s_1+r_1w_1),
\\
s_{1,t}=-\frac{4}{27}s_1(p_1s_1+r_1w_1),
\\
w_{1,t}=-\frac{4}{27}w_1(p_1s_1+r_1w_1),
\end{array}\right. \label{N1}
\end{eqnarray}
which has the solution
\begin{eqnarray}
q_{1}=C_1,\quad p_{1}=A_4e^{\frac{4}{27}(A_2+A_3)t},\quad r_{1}=\frac{1}{A_1}p_1,\quad
s_{1}=\frac{A_2}{A_4}e^{-\frac{4}{27}(A_2+A_3)t},
\quad
w_{1}=\frac{A_3}{r_1},
\label{N1s}
\end{eqnarray}
where $C_1$ and $A_1$, $\cdots$, $A_4$ are integration constants. Thus, the stationary single-peakon solution becomes
\begin{eqnarray}
\begin{array}{l}
u_{1}=A_4e^{\frac{4}{27}(A_2+A_3)t}e^{-\mid x-C_1\mid},\quad u_{2}=\frac{u_1}{A_1},
\\
v_{1}=\frac{A_2}{A_4}e^{-\frac{4}{27}(A_2+A_3)t}e^{-\mid x-C_1\mid},\quad v_{2}=\frac{A_1A_3}{A_2}v_1.
\end{array}
\label{ocpnp2}
\end{eqnarray}
See Figure \ref{F1} for the stationary single-peakon of the potentials $u_{1}(x,t)$ and $v_{1}(x,t)$ with $C_1=0$, $A_2=A_4=1$ and $A_3=2$.

\subsection*{Example 2.~~A new integrable four-component system with peakon solutions}
Choosing
$$H=-\frac{1}{9}[(u_{1}-u_{1,x})(v_{1}+v_{1,x})+(u_{2}-u_{2,x})(v_{2}+v_{2,x})],$$
equation (\ref{teq1}) is cast into
\begin{eqnarray}
\left\{\begin{array}{l}
m_{1,t}=(m_{1}H)_{x}+\frac{1}{9}m_1(u_{1}-u_{1,x})(v_{1}+v_{1,x})+\frac{1}{9}m_{2}(u_{1}-u_{1,x})(v_{2}+v_{2,x}),
\\
m_{2,t}=(m_{2}H)_{x}+\frac{1}{9}m_1(u_{2}-u_{2,x})(v_{1}+v_{1,x})+\frac{1}{9}m_{2}(u_{2}-u_{2,x})(v_{2}+v_{2,x}),
\\
n_{1,t}=(n_{1}H)_{x}-\frac{1}{9}n_1(u_{1}-u_{1,x})(v_{1}+v_{1,x})-\frac{1}{9}n_{2}(u_{2}-u_{2,x})(v_{1}+v_{1,x}),
\\
n_{2,t}=(n_{2}H)_{x}-\frac{1}{9}n_1(u_{1}-u_{1,x})(v_{2}+v_{2,x})-\frac{1}{9}n_{2}(u_{2}-u_{2,x})(v_{2}+v_{2,x}),
\\ m_{1}=u_{1}-u_{1,xx},\quad  m_{2}=u_{2}-u_{2,xx}, \quad n_{1}=v_{1}-v_{1,xx},\quad  n_{2}=v_{2}-v_{2,xx}.
\end{array}\right. \label{teq3}
\end{eqnarray}

Let us set
\begin{eqnarray}
\begin{split}
K=\frac{1}{9}\left( \begin{array}
{cccc} K_{11} & K_{12}  & K_{13} & K_{14} \\
K_{21} &  K_{22}  & K_{23} & K_{24} \\
K_{31} &  K_{32} & K_{33} & K_{34} \\
K_{41} &  K_{42} & K_{43} & K_{44} \\
\end{array} \right),
\label{JK2}
\end{split}
\end{eqnarray}
where
\begin{eqnarray}
\begin{split}
K_{11}=& \partial m_1\partial^{-1}m_1\partial-m_1\partial^{-1}m_1, ~~K_{12}=\partial m_1\partial^{-1}m_2\partial-m_2\partial^{-1}m_1,
\\ K_{13}=& \partial m_1\partial^{-1}n_1\partial+m_1\partial^{-1}n_1+m_2\partial^{-1}n_2, ~~K_{14}=\partial m_1\partial^{-1}n_2\partial,
\\ K_{21}=&-K_{12}^{\ast}=\partial m_2\partial^{-1}m_1\partial-m_1\partial^{-1}m_2, ~~K_{22}=\partial m_2\partial^{-1}m_2\partial-m_2\partial^{-1}m_2,
\\ K_{23}=& \partial m_{2}\partial^{-1} n_{1}\partial, ~~K_{24}=\partial m_2\partial^{-1}n_2\partial+m_1\partial^{-1}n_1+m_2\partial^{-1}n_2,
\\ K_{31}=&-K_{13}^{\ast}=\partial n_1\partial^{-1}m_1\partial+n_1\partial^{-1}m_1+n_2\partial^{-1}m_2, ~~K_{32}=-K_{23}^{\ast}=\partial n_1\partial^{-1} m_2\partial,
\\ K_{33}=&\partial n_{1}\partial^{-1} n_{1}\partial-n_{1}\partial^{-1} n_{1}, ~~K_{34}=\partial n_{1}\partial^{-1} n_{2}\partial-n_{2}\partial^{-1} n_{1},
\\ K_{41}=&-K_{14}^{\ast}=\partial n_2\partial^{-1} m_1\partial, ~~K_{42}=-K_{24}^{\ast}=\partial n_2\partial^{-1}m_2\partial+n_1\partial^{-1}m_1+n_2\partial^{-1}m_2,
\\ K_{43}=&-K_{34}^{\ast}=\partial n_2\partial^{-1}n_1\partial-n_1\partial^{-1}n_2, ~~K_{44}=\partial n_2\partial^{-1}n_2\partial-n_2\partial^{-1}n_2.
\end{split}
\label{K2}
\end{eqnarray}
Direct calculations yield that
\begin{proposition}
Equation (\ref{teq3}) can be rewritten in the following Hamiltonian form
\begin{eqnarray}
\left(m_{1,t}, ~m_{2,t}, ~n_{1,t}, ~n_{2,t}\right)^{T}
=K \left(\frac{\delta H_1}{\delta m_{1}},~\frac{\delta H_1}{\delta  m_{2}},~\frac{\delta H_1}{\delta n_{1}},~\frac{\delta H_1}{\delta n_{2}}\right)^{T},
\label{BH2}
\end{eqnarray}
where $K$ are given by (\ref{JK2}), and
\begin{eqnarray}
\begin{split}
H_1&=\int_{-\infty}^{+\infty}[(u_{1,x}-u_{1})n_{1}+(u_{2,x}-u_{2})n_{2}]dx.
\end{split}
\label{H2}
\end{eqnarray}
\end{proposition}

We believe that the equation (\ref{teq3}) could be cast into a bi-Hamiltonian system. But we didn't
find another Hamiltonian operator yet that is compatible with the Hamiltonian operator (\ref{JK2}).

Suppose $N$-peakon solution of (\ref{teq3}) is expressed also in the form of (\ref{NP2}). Then,
we obtain the $N$-peakon dynamical system of (\ref{teq3}):
\begin{eqnarray}
\begin{split}
q_{j,t}=&\frac{1}{9}\{-\frac{1}{3}(p_js_j+r_jw_j)
         +\sum_{i,k=1}^N \left(p_is_k+r_iw_k\right)\left(sgn(q_j-q_i)-sgn(q_j-q_k)\right)e^{ -\mid q_j-q_i\mid-\mid q_j-q_k\mid}
         \\&+\sum_{i,k=1}^N \left(p_is_k+r_iw_k\right)\left(1-sgn(q_j-q_i)sgn(q_j-q_k)\right)e^{ -\mid q_j-q_i\mid-\mid q_j-q_k\mid}\},\\
p_{j,t}=&\frac{1}{9}\{-\frac{1}{3}p_j(p_js_j+r_jw_j)
         +\sum_{i,k=1}^N \left(p_jp_is_k+r_jp_iw_k\right)\left(sgn(q_j-q_i)-sgn(q_j-q_k)\right)e^{ -\mid q_j-q_i\mid-\mid q_j-q_k\mid}
         \\&+\sum_{i,k=1}^N \left(p_jp_is_k+r_jp_iw_k\right)\left(1-sgn(q_j-q_i)sgn(q_j-q_k)\right)e^{ -\mid q_j-q_i\mid-\mid q_j-q_k\mid}\},\\
r_{j,t}=&\frac{1}{9}\{-\frac{1}{3}r_j(p_js_j+r_jw_j)
         +\sum_{i,k=1}^N \left(r_jr_iw_k+p_jr_is_k\right)\left(sgn(q_j-q_i)-sgn(q_j-q_k)\right)e^{ -\mid q_j-q_i\mid-\mid q_j-q_k\mid}
         \\&+\sum_{i,k=1}^N \left(r_jr_iw_k+p_jr_is_k\right)\left(1-sgn(q_j-q_i)sgn(q_j-q_k)\right)e^{ -\mid q_j-q_i\mid-\mid q_j-q_k\mid}\},\\
s_{j,t}=&\frac{1}{9}\{\frac{1}{3}s_j(p_js_j+r_jw_j)
         -\sum_{i,k=1}^N \left(w_jr_is_k+s_jp_is_k\right)\left(sgn(q_j-q_i)-sgn(q_j-q_k)\right)e^{ -\mid q_j-q_i\mid-\mid q_j-q_k\mid}
         \\&+\sum_{i,k=1}^N \left(s_jp_is_k+w_jr_is_k\right)\left(sgn(q_j-q_i)sgn(q_j-q_k)-1\right)e^{ -\mid q_j-q_i\mid-\mid q_j-q_k\mid}\},\\
w_{j,t}=&\frac{1}{9}\{\frac{1}{3}w_j(p_js_j+r_jw_j)
         -\sum_{i,k=1}^N \left(s_jp_iw_k+w_jr_iw_k\right)\left(sgn(q_j-q_i)-sgn(q_j-q_k)\right)e^{ -\mid q_j-q_i\mid-\mid q_j-q_k\mid}
         \\&+\sum_{i,k=1}^N \left(w_jr_iw_k+s_jp_iw_k\right)\left(sgn(q_j-q_i)sgn(q_j-q_k)-1\right)e^{ -\mid q_j-q_i\mid-\mid q_j-q_k\mid}\}.
\end{split}
\label{dNcp3}
\end{eqnarray}

For $N=1$, (\ref{dNcp3}) becomes
\begin{eqnarray}
\left\{\begin{array}{l}
q_{1,t}=\frac{2}{27}(p_1s_1+r_1w_1),
\\
p_{1,t}=\frac{2}{27}p_1(p_1s_1+r_1w_1),
\\
r_{1,t}=\frac{2}{27}r_1(p_1s_1+r_1w_1),
\\
s_{1,t}=-\frac{2}{27}s_1(p_1s_1+r_1w_1),
\\
w_{1,t}=-\frac{2}{27}w_1(p_1s_1+r_1w_1).
\end{array}\right. \label{N2}
\end{eqnarray}
We may solve this equation as
\begin{eqnarray}
q_{1}=\frac{2}{27}(A_2+A_3)t+A_4,~p_{1}=A_5e^{\frac{2}{27}(A_2+A_3)t},~r_{1}=\frac{1}{A_1}p_1,~
s_{1}=\frac{A_2}{A_5}e^{-\frac{2}{27}(A_2+A_3)t},~w_{1}=\frac{A_3}{r_1},
\label{N2s}
\end{eqnarray}
where $A_1$, $\cdots$, $A_5$ are integration constants.
See Figure \ref{F2} for the single-peakon of the potentials $u_{1}(x,t)$ and $v_{1}(x,t)$ with $A_4=0$, $A_2=A_5=1$ and $A_3=2$.

\section {Conclusions and discussions}
In our paper, we propose a multi-component generalization of the Camassa-Holm equation, and provide its Lax representation and infinitely many conservation laws.
This system contains an arbitrary smooth function $H$, thus it is actually a large class of multi-component peakon equations.
Due to the presence of the arbitrary function, we do not expect that the system is bi-Hamiltonian in the general case.
But we show it is possible to find the bi-Hamiltonian structures for the special choices of $H$.
In particular, we study the peakon solutions of this system in the case $N=2$, and obtain a new integrable system which admits stationary peakon solutions.

As mentioned above, Li, Liu and Popowicz proposed a four-component system which also contains an arbitrary function \cite{LLP}.
In contrast with the usual soliton equations, the peakon equations with arbitrary functions seem to be unusual.
We believe that there are much investigations deserved to do for both our generalized peakon system and Li-Liu-Popowicz's system.
The following topics seem to be interesting:
\\ (1) Is there a gauge transformation that can be applied to the Lax pair to remove the arbitrary function $H$?
\\ (2) Can the inverse scattering transforms be applied to solve the systems in general?
\\ (3) Do there exist infinitely many commuting symmetries for the systems?

\section*{ACKNOWLEDGMENTS}

This work was partially supported by the National Natural Science Foundation of China (Grant Nos. 11301229, 11271168, and 11171295), the Natural Science Foundation of the Jiangsu Province (Grant No. BK20130224), the Natural Science Foundation of the Jiangsu Higher Education Institutions of China (Grant No. 13KJB110009), and the China state administration of foreign experts affairs system under the affiliation of China University of Mining and Technology.

\vspace{1cm}
\small{

}
\end{document}